\documentclass[prb]{revtex4}

\usepackage{graphicx}
\usepackage{dcolumn}
\usepackage{bm}

\begin{document}

\preprint{PRB}

\title
{ Anisotropic susceptibilities of thin ferromagnetic films within
many-body Green's function theory }


\author{P. Fr\"obrich}
\altaffiliation[Also at ]{Institut f\"ur Theoretische Physik,
Freie Universit\"at Berlin,
Arnimallee 14, D-14195 Berlin, Germany.}

\email{froebrich@hmi.de}

\author{P.J. Kuntz}
\email{kuntz@hmi.de}
\affiliation{Hahn-Meitner-Institut Berlin,
Glienicker Stra{\ss}e 100, D-14109 Berlin,
Germany}


\date{\today}

\def\K{\mathord{\cal K}}

\def\la{\langle}

\def\ra{\rangle}

\def\ltsim{\mathop{\,<\kern-1.05em\lower1.ex\hbox{$\sim$}\,}}

\def\gtsim{\mathop{\,>\kern-1.05em\lower1.ex\hbox{$\sim$}\,}}

\begin{abstract}
Transverse and parallel static susceptibilities of in-plane uniaxial
anisotropic
ferromagnetic films are calculated using a Heisenberg model within the
framework of many-body Green's function theory. The importance of
collective
magnetic excitations, in particular in the paramagnetic regime,
is demonstrated by
comparing with mean field calculations. The paper extends the work of
Jensen et al. \cite{Jens03} on the monolayer with spin $1/2$ to the
multilayer case with arbitrary spin.
\end{abstract}

\pacs{75.10.Jm,
75.70.Ak,
75.30Ds
}




\maketitle

\subsection*{1. Introduction}
In a recent paper, Jensen et al. \cite{Jens03} reported on the measurement of
the magnetic susceptibility of a bilayer Co film with an in-plane uniaxial
anisotropy. On the basis of a Heisenberg model they determined the isotropic
exchange interaction and the magnetic anisotropy within the
framework of a Green's
function theory by fitting the interaction
parameters of the theory to the measured susceptibilities along the easy and
hard axes in the paramagnetic regime, assuming a spin $S=1/2$.
In this paper, we generalize the
theoretical treatment to the multilayer case and to spins
S$>$1/2. We organize the paper as follows. In Section 2 we explain the model
and the Green's function formalism for its solution. Section 3 displays the
numerical results. In the final Section 4 we summarize the results and
present our conclusions.

\subsection*{2. The model and the Green's function formalism}

We consider a Hamiltonian consisting of a ferromagnetic isotropic
Heisenberg exchange
interaction with strength ($J_{kl}>0$) between nearest neighbour lattice sites,
a uniaxial in-plane exchange
anisotropy in the z-direction with strength ($D_{kl}>0$),
and an external magnetic
field ${\bf B}=(B^x,0,B^z)$ confined to the film plane:
\begin{eqnarray}
{\cal H}=&-&\frac{1}{2}\sum_{<kl>}J_{kl}(S_k^-S_l^++S_k^zS_l^z)
-\frac{1}{2}\sum_{<kl>}D_{kl}S_k^zS_l^z
\nonumber\\
&-&\sum_k\Big(B^x\frac{1}{2}(S_k^++S_k^-)+B^zS_k^z\Big).
\label{1}
\end{eqnarray}
Here the notation $S_k^{\pm}=S_k^x\pm iS_k^y$  is
introduced, where $k$ and $l$ are lattice site indices and $<kl>$ indicates
summation over nearest neighbours only. In keeping with reference
\cite{Jens03},
we do not consider the dipole coupling, since it is almost isotropic for
the in-plane situation. We note, however, that
the formalism is capable of handling this coupling, if so desired, and
we refer the reader to references \cite{FJKE00,FK03}, where we
treated the reorientation of the magnetization of ferromagnetic films with
anisotropies normal to the film plane. Also in keeping with reference
\cite{Jens03}, we choose
the exchange anisotropy over the single-ion anisotropy, which has the
advantage of being simpler to
handle in the Green's function theory.
Moreover, we have shown in \cite{FK03} that
once the strength of the exchange anisotropy is fitted appropriately,
the magnetization as a function of the
temperature and film thickness behaves very similar to that calculated from
the single-ion anisotropy, which may appear somewhat surprising, since the
anisotropies originate from very different physical mechanisms.
In this paper we shall restrict ourselves to a simple cubic lattice.

 In order to generalize the treatment of \cite{Jens03} to
general spin $S$, we need the following Green's functions
\begin{equation}
G_{ij,\eta}^{\alpha,mn}(\omega)=\la\la
S_i^\alpha;(S_j^z)^m(S_j^-)^n\ra\ra_{\omega,\eta}\ ,
\label{2}
\end{equation}
where $\alpha=(+,-,z)$ takes care of all directions in space, $\eta=\pm 1$
refers to the anticommutator or commutator Green's functions, respectively, and
$n\geq 1, m\geq 0$ are positive integers, necessary for dealing with higher
spin values $S$. For $n=1$ and $m=0$ we will recover the equations of
\cite{Jens03}.

The exact equations of motion for the Green's functions are
\begin{equation}
\omega G_{ij,\eta}^{\alpha,mn}(\omega)=A_{ij,\eta}^{\alpha,mn}+\la\la
[S_i^\alpha,{\cal H}]_-;(S_j^z)^m(S_j^-)^n\ra\ra_{\omega,\eta}
\label{3}
\end{equation}
with the inhomogeneities
\begin{equation}
A_{ij,\eta}^{\alpha,mn}=\la[S_i^\alpha,(S_j^z)^m(S_j^-)^n]_{\eta}\ra,
\label{4}
\end{equation}
where $\la ...\ra=Tr(...e^{-\beta{\cal H}})/Tr(e^{-\beta{\cal H}})$ denotes the
thermodynamic expectation value.

After solving these equations, the components of the magnetization can be
determined from the Green's functions via the spectral theorem.
A closed system of equations is achieved by
decoupling the higher-order Green's functions on the right hand sides.
For the exchange-interaction and exchange-anisotropy terms, we use a
generalized Tyablikov- ( RPA-) decoupling
\begin{equation}
\la\la S_i^\alpha S_k^\beta;(S_j^z)^m(S_j^-)^n\ra\ra_\eta \simeq\la
S_i^\alpha\ra
G_{kj,\eta}^{\beta,mn}+\la S_k^\beta\ra G_{ij,\eta}^{\alpha,mn} .
\label{5}
\end{equation}
We do not try to do better than RPA because we have shown in earlier
work\cite{EFJK99,HFKTJ02}, by comparing RPA with ``exact''
Quantum Monte Carlo calculations that the former is quite a good
approximation in simple cases.

We now proceed with the formulation of the theory for the multilayer case.
After a Fourier transform to momentum
space, we obtain, for a film with $N$ layers,
$3N$ equations of motion for a $3N$-dimensional Green's function vector ${\bf
G}^{mn}$:
\begin{equation}
(\omega{\bf 1}-{\bf \Gamma}){\bf G}^{mn}_\eta={\bf A}^{mn}_\eta,
\label{6}
\end{equation}
where ${\bf 1}$ is the
$3N\times 3N$ unit matrix. The Green's function vectors and inhomogeneity
vectors each
consist of $N$  three-dimensional subvectors which are characterized by the
layer indices $i$ and $j$

\begin{equation}
{\bf G}_{ij,\eta}^{mn}({\bf{k},\omega})\  =
\left( \begin{array}{c}
G_{ij,\eta}^{+,mn}({\bf{k}},\omega) \\ G_{ij,\eta}^{-,mn}({\bf{k}},\omega)  \\
G_{ij,\eta}^{z,mn}({\bf{k}},\omega)
\end{array} \right), \hspace{0.5cm}
{\bf A}_{ij,\eta}^{mn} {=}
 \left( \begin{array}{c} A_{ij,\eta}^{+,mn} \\ A_{ij,\eta}^{-,mn} \\
A_{ij,\eta}^{z,mn} \end{array} \right) \;.
\label{7} \end{equation}

The equations of motion are then expressed in terms of these layer vectors and
the $3\times 3 $ submatrices ${\bf \Gamma}_{ij}$ of the $3N\times 3N$
matrix ${\bf\Gamma}$
\begin{eqnarray}
& &\left[ \omega {\bf 1}-\left( \begin{array}{cccc}
{\bf\Gamma}_{11} & {\bf\Gamma}_{12} & \ldots & {\bf\Gamma}_{1N} \\
{\bf\Gamma}_{21} & {\bf\Gamma}_{22} & \ldots & {\bf\Gamma}_{2N} \\
\ldots & \ldots & \ldots & \ldots \\
{\bf\Gamma}_{N1} & {\bf\Gamma}_{N2} & \ldots & {\bf\Gamma}_{NN}
\end{array}\right)\right]\left[ \begin{array}{c}
{\bf G}_{1j,\eta} \\ {\bf G}_{2j,\eta} \\ \ldots \\ {\bf G}_{Nj,\eta}
\end{array} \right]\nonumber\\&=&\left[ \begin{array}{c}
{\bf A}_{1j,\eta}\delta_{1j} \\ {\bf A}_{2j,\eta}\delta_{2j} \\ \ldots \\
{\bf A}_{Nj,\eta}\delta_{Nj} \end{array}
\right] \;, \hspace{0.3cm} j=1,...,N\;.
\label{8}
\end{eqnarray}
When performing the decouplings according to equation (\ref{5}), the  ${\bf
\Gamma}$-matrix reduces to a band matrix
with zeros in the ${\bf \Gamma}_{ij}$ sub-matrices, when $j>i+1$ and $j<i-1$.
The  diagonal sub-matrices ${\bf \Gamma}_{ii}$ are of size $3\times 3$
and have the form
\begin{equation}
 {\bf \Gamma}_{ii}= \left( \begin{array}
{@{\hspace*{3mm}}c@{\hspace*{5mm}}c@{\hspace*{5mm}}c@{\hspace*{3mm}}}
\;\;\;H^z_i & 0 & -H^x_i \\ 0 & -H^z_i & \;\;\;H^x_i \\
-\frac{1}{2}\tilde{H}^x_i & \;\frac{1}{2}\tilde{H}^x_i & 0
\end{array} \right)
\ . \label{9}
\end{equation}
where
\begin{eqnarray}
H^z_i&=&Z_i+\la S_i^z\ra J_{ii}(q-\gamma_{\bf k})\ ,
\nonumber\\
Z_i&=&B^z
+D_{ii}q\la S^z_i\ra
+(J_{i,i+1}+D_{i,i+1})\la S_{i+1}^{z}\ra
\nonumber\\
& &+(J_{i,i-1}+D_{i,i-1})\la
S_{i-1}^{z}\ra\ ,\\
\tilde{H}^x_i&=&B^x+\la S_i^x\ra J_{ii}(q-\gamma_{\bf
k})
+J_{i,i+1}\la S_{i+1}^{x}\ra+J_{i,i-1}\la
S_{i-1}^{x}\ra \ ,
\nonumber\\
H^x_i&=&\tilde{H}^x_i-\la S_i^x\ra D_{ii}\gamma_{\bf k} \ .
\nonumber
\label{10}
\end{eqnarray}
For a square lattice and a lattice constant taken to be unity,  $\gamma_{\bf
k}=2(\cos k_x+\cos k_z)$, and $q=4$ is the
number of intra-layer nearest neighbours. The mean field (MFT) results, which
we use later for comparison with Green's function theory, are obtained by
putting $\gamma_{\bf k}=0$; i.e. only the number of nearest neighbour counts,
whereas RPA introduces a momentum dependence on the lattice under
consideration.

Note that because
the momentum dependence in $H_i^x$ stems from the exchange anisotropy,
 $\tilde{H}_i^x\neq H_i^x$, which prevents
a naive extension of the formalism of reference \cite{FJKE00}.

The $3\times 3$
off-diagonal sub-matrices ${\bf \Gamma}_{ij}$ for $j= i\pm 1$ are of the
form
\begin{equation}
 {\bf \Gamma}_{ij} = \left( \begin{array}
{@{\hspace*{3mm}}c@{\hspace*{5mm}}c@{\hspace*{5mm}}c@{\hspace*{3mm}}}
-J_{ij}\la S_i^z\ra & 0 & \;\;\;(J_{ij}+D_{ij})\la S_i^x\ra \\
0 & \;\;J_{ij}\la S_i^z\ra & -(J_{ij}+D_{ij)}\la S_i^x\ra \\
\frac{1}{2}J_{ij}\la S_i^x\ra &
-\frac{1}{2}J_{ij}\la S_i^x\ra & 0 \end{array} \right) \;.
\label{11}
\end{equation}

The treatment of multilayers is only practicable when
using the eigenvector method
developed in reference \cite{FJKE00}.
The essential features are as follows.
One starts with a transformation, which diagonalizes the ${\bf \Gamma}$-matrix
of equation (\ref{6})
\begin{equation}
{\bf L\Gamma R}={\bf \Omega},
\label{12}
\end{equation}
where ${\bf \Omega}$ is a diagonal matrix with eigenvalues $\omega_{\tau}$
($\tau=1,..., 3N$), and the transformation matrix {\bf R} and its inverse
${\bf R}^{-1}={\bf L}$ are obtained from the right eigenvectors of
${\bf \Gamma}$
as columns and from the left eigenvectors as rows, respectively. These matrices
are normalized to unity: {\bf RL}={\bf LR}={\bf 1}.

Multiplying the equation of motion (\ref{6}) from the left by {\bf L} and
inserting {\bf 1}={\bf RL} one finds
\begin{equation}
(\omega{\bf 1}-{\bf \Omega}){\bf L}{\bf G}_\eta^{mn}={\bf LA}_\eta^{mn}.
\label{13}
\end{equation}
Defining ${\cal G}_\eta^{mn}={\bf LG}_\eta^{mn}$
and ${\cal A}_\eta^{mn}={\bf LA}_\eta^{mn}$ one obtains
\begin{equation}
(\omega {\bf 1}-{\bf \Omega}){\cal G}_\eta^{mn}={\cal A}_\eta^{mn}.
\label{14}
\end{equation}
${\cal G}_\eta^{mn}$ is a vector of new Green's functions, each component
$\tau$ of which has but a single pole
\begin{equation}
{\cal G}_\eta^{mn,\tau}=\frac{{\cal A}_\eta^{mn,\tau}}{\omega-\omega_\tau}\ .
\label{15}
\end{equation}
This is the important point because it allows application of the spectral
theorem, e.g. \cite{GHE01}, to
each component separately. We obtain for the component $\tau$ of correlation
vector ${\cal C}^{mn}={\bf L}{\bf C}^{mn}$
( where ${\bf C}^{mn}=\la (S^z)^m(S^-)^nS^\alpha\ra$ with $(\alpha=+,-,z)$)
\begin{equation}
{\cal C}^{mn,\tau}=\frac{{\cal
A}_{\eta}^{mn,\tau}}{e^{\beta\omega_\tau}+\eta}+\frac{1}{2}(1-\eta)\frac{1}{2}
\lim_{\omega\rightarrow 0}\omega \frac{{\cal
A}_{\eta=+1}^{mn,\tau}}{\omega-\omega_\tau} .
\label{16}
\end{equation}
We emphasize that when ($\eta=-1$), the second term of this
equation, which is due to the anticommutator Green's function, has to be taken
into account. This term occurs for $\omega_\tau=0$ and can be simplified by
using the relation between anticommutator and commutator
\begin{equation}
{\cal A}_{\eta=+1}^{mn,0}={\cal A}^{mn,0}_{\eta=-1}+2{\cal
C}^{mn,0}={\bf L}_0(A_{\eta=-1}^{mn}+2{\bf C}^{mn}),
\label{17}
\end{equation}
where the index $\tau=0$ refers to the eigenvector with $\omega_\tau=0$.

The term ${\bf L}_0A_{\eta=-1}^{mn}=0$ vanishes due to the fact that the
commutator Green's function has to be regular at the origin
\begin{equation}
\lim_{\omega\rightarrow 0}\omega G_{\eta=-1}^{\alpha,mn}=0,
\label{18}
\end{equation}
which leads to the regularity conditions:
\begin{equation}
\tilde{H}^xA_{\eta=-1}^{+,mn}+\tilde{H}^xA_{\eta=-1}^{-,mn}+2H^zA_{\eta=-1}^{z,
m n } =0.
\label{19}
\end{equation}
For details, see reference \cite{FJKE00}.

This is equivalent to
\begin{equation}
{\bf L}_0A_{\eta=-1}^{mn}=0,
\label{20}
\end{equation}
because the left eigenvector of the ${\bf \Gamma}$-matrix with eigenvector zero
has the structure (see also equation (\ref{27}) below)
\begin{equation}
{\bf L}_0\propto (\tilde{H}^x, \tilde{H}^x, 2H^z).
\label{21}
\end{equation}
For more details concerning the use of the regularity conditions, see refs.
\cite{FJKE00,FK03}.

The equations for the correlations are then obtained by multiplying equation
(\ref{16})  from the left with ${\bf R}$ and using equation (\ref{20});
i.e.
\begin{equation}
{\bf C}={\bf R}{\bf {\cal E}}{\bf L}{\bf A}+{\bf R}_0{\bf L}_0{\bf C},
\label{22}
\end{equation}
where ${\bf {\cal E}}$ is a diagonal matrix with matrix elements
${\cal E}_{ij}=\delta_{ij}(e^{\beta\omega_i}-1)^{-1}$ for eigenvalues
$\omega_i\neq 0$, and $0$ for eigenvalues $\omega_i=0$.

A problem associated with this equation is that the
exchange anisotropy introduces a momentum dependence into
the projection operator ${\bf R}_0{\bf L}_0$.
Consequently, when the
Fourier transform to real space is performed, the projector cannot be
taken out of the integral
as is possible in the case of the Anderson Callen decoupling of the
single-ion anisotropy  \cite{FJKE00}, where the projector turns out to be
momentum independent.
The solution is to eliminate one component of the projector by a
transformation
of equation (\ref{22}), which is sufficient to  establish the integral
equations of the eigenvector method.

The adequate transformation  is found to be
\begin{equation}
{\bf T}^{-1}=\frac{1}{2}\left(\begin{array}{ccc}
1 & 1 & 0 \\
-1& 1 & 0  \\
0 & 0 & 2
\end{array}\right)\ \ \ \ \ \ \
{\bf T}=\left(\begin{array}{ccc}
1 & -1 & 0\\
1 & 1 & 0 \\
0 & 0& 1
\end{array}\right)
\label{23}
\end{equation}
with ${\bf T}^{-1}{\bf T}={\bf 1}$.

Applying this transformation to equation (\ref{22}) considered as a monolayer
problem
\begin{equation}
{\bf T}^{-1}{\bf C}={\bf T}^{-1}{\bf R{\cal E}LTT}^{-1}{\bf A}+{\bf
T}^{-1}{\bf R}_0{\bf L}_0{\bf TT}^{-1}{\bf C}
\label{24}
\end{equation}
transforms the
second component of the vector
${\bf T}^{-1}{\bf R}_0{\bf L}_0{\bf TT}^{-1}{\bf C}$ to zero.
This can be seen when inserting the explicit expressions
for the eigenvalues and eigenvectors for the monolayer.
The eigenvalues of the ${\bf \Gamma}$-matrix in this case are
\begin{equation}
\omega_1=0;\ \ \omega_{2,3}=\pm\varepsilon_{\bf
k}=\pm\sqrt{H^zH^z+\tilde{H^x}H^x},
\label{25}
\end{equation}
the right eigenvectors are the columns of the matrix
\begin{equation}
{\bf R}=\left(\begin{array}{ccc}
\frac{H^x}{H^z} &\frac{-(\epsilon_{\bf
k}+H^z)}{\tilde{H}^x}&\frac{(\epsilon_{\bf k}-H^z)}{\tilde{H}^x}\\
\frac{H^x}{H^z} &\frac{(\epsilon_{\bf
k}-H^z)}{\tilde{H}^x}&\frac{-(\epsilon_{\bf k}+H^z)}{\tilde{H}^x}\\
1&1&1
\end{array}\right)\ ,
\label{26}
\end{equation}
and the left eigenvectors are the rows of the matrix
\begin{equation}
{\bf L}=\frac{1}{4\epsilon_{\bf k}^2}\left(
\begin{array}{ccc}
2\tilde{H^x}H^z&2\tilde{H^x}H^z&4H^zH^z\\
-(\epsilon_{\bf k}+H^z)\tilde{H}^x&(\epsilon_{\bf
k}-H^z)\tilde{H}^x&2H^x\tilde{H}^x\\
(\epsilon_{\bf k}-H^z)\tilde{H}^x&-(\epsilon_{\bf
k}+H^z)\tilde{H}^x&2H^x\tilde{H}^x
\end{array}\right)\ .
\label{27}
\end{equation}

The second row of the
transformed equation (\ref{24}), together with the regularity
conditions (\ref{19}), leads to one integral equation for
the correlations for each ($m,n$)-pair.

The eigenvector method can
immediately be generalized to the case of $N$ layers by
transforming equation (\ref{22}) (extended to
$3N$-dimensions) with a $3N\times 3N$-matrix  ${\bf T^{-1}}$ having the
$3\times 3$ $T^{-1}$ sub-matrices  (\ref{23})  on the diagonal.

Before showing numerical results for $S>1/2$ and for multilayers we derive
the equations for the monolayer and $S=1/2$ of reference \cite{Jens03} from
equation (\ref{24}).

Using equations (\ref{26},\ref{27}), one obtains from the second row of
equation
(\ref{24}) for the monolayer and for general spin
\begin{eqnarray}
C^{+,mn}-C^{-,mn}+\frac{1}{2}(A^{+,mn}_{\eta=-1}-A^{-,mn}_{\eta=-1})
=
\nonumber\\
\frac{1}{2}(A^{+,mn}_{\eta=-1}+A^{-,mn}_{\eta=-1})\frac{1}{N}\sum_{\bf
k}\frac{\varepsilon_{\bf k}}{H^z}\coth({\beta \varepsilon_{\bf k}/2})
\label{28}
\end{eqnarray}
and from the first row
\begin{eqnarray}
& &C^{+,mn}-C^{-,mn}+\frac{1}{2}(A^{+,mn}_{\eta=-1}+A^{-,mn}_{\eta=-1})=\\
\nonumber & &\frac{2H^x}{H^z}C^{z,mn}+A^{z,mn}_{\eta=-1}\frac{H^x}{H^z}
\\ \nonumber& &
+\frac{1}{2}(A^{+,mn}_{\eta=-1}-A^{-,mn}_{\eta=-1})\frac{1}{N}
\sum_{\bf k}\frac{\varepsilon_{\bf k}}{H^z}\coth({\beta \varepsilon_{\bf
k}/2}).
\label{29}
\end{eqnarray}
The equations for $S=1/2$ are obtained from these equations   for $n=1,
m=0$, i.e. $C^{-,01}=\la S^-S^-\ra=0$, $C^{+,01}=\la S^-S^+\ra=1/2-\la S^z\ra$,
$A^{-,01}_{\eta=-1}=0$, $A^{+,01}_{\eta=-1}=2\la S^z\ra$,
$C^{z,01}=\la S^-S^z\ra=1/2\la S^x\ra$, and from the regularity condition
(\ref{19}) $\la S^z\ra=(H^z/\tilde{H^x})\la S^x\ra$:
\begin{equation}
\frac{1}{2}=\la S^z\ra \frac{1}{N}\sum_{\bf k}\frac{\varepsilon_{\bf
k}}{H^z}\coth({\beta\varepsilon_{\bf k}/2}),
\label{30}
\end{equation}
and
\begin{equation}
\frac{1}{2}=\la S^x\ra \frac{1}{N}\sum_{\bf k}\frac{\varepsilon_{\bf
k}}{\tilde{H^x}}\coth({\beta\varepsilon_{\bf k}/2}) ,
\label{31}
\end{equation}
which are the equations used in ref.\cite{Jens03}.

For $S>1/2$ and for the multilayer case, the eigenvectors have to be calculated
numerically and the system of integral equations obtained from the
Fourier transform of equations (\ref{24}) to real space has to be solved
self-consistently, which is done by the curve-following method described
in detail in \cite{FKS02}.

\subsection*{3. Numerical results}
\subsection*{3.1. The monolayer with arbitrary spin ${\bf S}$}
We start by showing results for the magnetizations $\la S^z\ra$ and
$\la S^x\ra$ with respect to the easy and hard axes of a monolayer with various
spin values $S$. It turns out that one obtains fairly universal curves when
scaling the parameters of the model as
$\tilde{J}/S(S+1)=J, \tilde{D}/S(S+1)=D$, and
$\tilde{B}^{x(z)}/S=B^{x(z)}$. In the present paper we
restrict ourselves to
the case where the exchange interaction and exchange anisotropy parameters are
the same for all layers and interlayer couplings. The program is, however,
written in such a way that different parameters can be easily introduced.

Using the scaled variables, the Curie temperatures $T_C(S)$ collapse to
a single universal value
(the same for each spin $S$) both in mean field theory (MFT) and in the
random phase approximation (RPA).
\begin{figure}[htb]
\begin{center}
\protect
\includegraphics*[bb = 80  85 520 410,
angle=-0,clip=true,width=9cm]{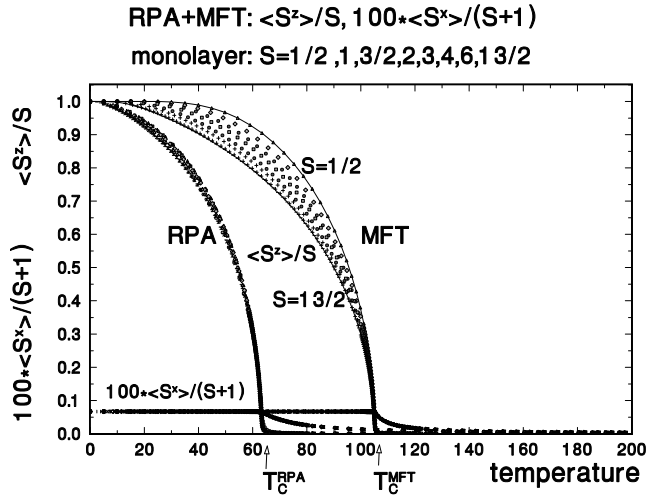}
\protect
\caption{The `universal' magnetizations $\la S^z\ra/S$ and
$100*\la S^x\ra/(S+1)$ of an
anisotropic ferromagnetic Heisenberg monolayer for a square lattice are shown
as functions of the temperature for $S=1/2,1,3/2,2,3,4,6,13/2$. Comparison
is made between Green's function (RPA) and mean field (MFT) calculations
using the exchange interaction $\tilde{J}=75$, the exchange
anisotropy $\tilde{D}=3.75$ (corresponding to $J=100, D=5$ of the S=1/2 case of
Ref.\cite{Jens03}) and small magnetic fields $\tilde{B}^x=\tilde{B}^z=0.01$. }
\label{fig1} \end{center}
\end{figure}

This is shown in Fig.1, where MFT and RPA results for
$\la S^z\ra/S$ and $100\times \la S^x\ra/(S+1)$ are displayed as functions of
the temperature for the spin values $S=1/2,1,3/2,2,3,4,6,13/2$. For the
exchange interaction  and exchange anisotropy, we use the same parameters as in
Ref. \cite{Jens03}, which is the S=1/2 monolayer case in our investigations.
The calculations for Fig.1
are done with small fields ($\tilde{B}^x=\tilde{B}^z=0.01$); this stabilizes
the
numerical algorithm. Whereas $\la S^z\ra/S$ in RPA is universal over the whole
temperature
range, the corresponding MFT curves as function of the temperature split
somewhat, reaching a saturation for large spin values S (approaching the
classical limit). The curves $100\times\la S^x\ra/(S+1)$ have the same
universal value in MFT and RPA for $T<T_C^{RPA(MFT)}$ (Because of the very
small $B^x$ field we introduced the factor 100 to make the curves visible).
The reason why the values for $\la S^x\ra/(S+1)$ coincide below the Curie
temperature is that the magnetization in $x$-direction depends only on the
number of nearest neighbours. This can be understood from equation (\ref{19})
from which one obtains with ($n=1,m=0$)
\begin{equation}
\la S^x\ra=\lim_{B^z\rightarrow
0}\frac{\tilde{H^x}}{H^z}=\frac{B^x}{Dq}=
\frac{\tilde{B}^xS(S+1)}{\tilde{D}qS}.
\label{32}
\end{equation}
This explains the universality of $\la S^x\ra/(S+1)=\tilde{B}^x/(\tilde{D}q)$,
where $q=4$ is the number of nearest neighbours for the square monolayer.
The fact that the universal Curie temperature $T_C^{RPA}(S)$ is only about one
half of $T_C^{MFT}(S)$ for the monolayer is due to the action of collective
excitations (magnons=spin waves), which are completely absent in MFT.

Spin waves also have significant effects on the susceptibilities (in particular
in the paramagnetic regime $T>T_C$) with respect to the easy ($\chi_{zz}$) and
hard ($\chi_{xx}$) axes. The susceptibilities are calculated as differential
quotients
\begin{eqnarray}
\chi_{zz}&=&\Big(\la S^z(B^z)\ra-\la S^z(0)\ra\Big)/B^z\nonumber\\
\chi_{xx}&=&\Big(\la S^x(B^x)\ra-\la S^x(0)\ra\Big)/B^x,
\end{eqnarray}
where the use of $B^z=B^x=0.01/S$ turns out to be small enough to get
good numerical estimates of the quotients; smaller fields would only be
necessary to get better estimates close to the divergence of
$\chi_{zz}$
at $T_C$; however, the errors in the inverse susceptibilities $\chi_{zz}^{-1}$
and $\chi_{xx}^{-1}$ at this point are not noticeable in the figures.
The inverse susceptibilities as functions of the temperature can be
brought into near coinicidence with a single universal curve if they are
multiplied with a factor $S(S+1)$, especially in the paramagnetic regime.
\begin{figure}[htb]
\begin{center}
\protect
\includegraphics*[bb = 80  90 520 420,
angle=-0,clip=true,width=9cm]{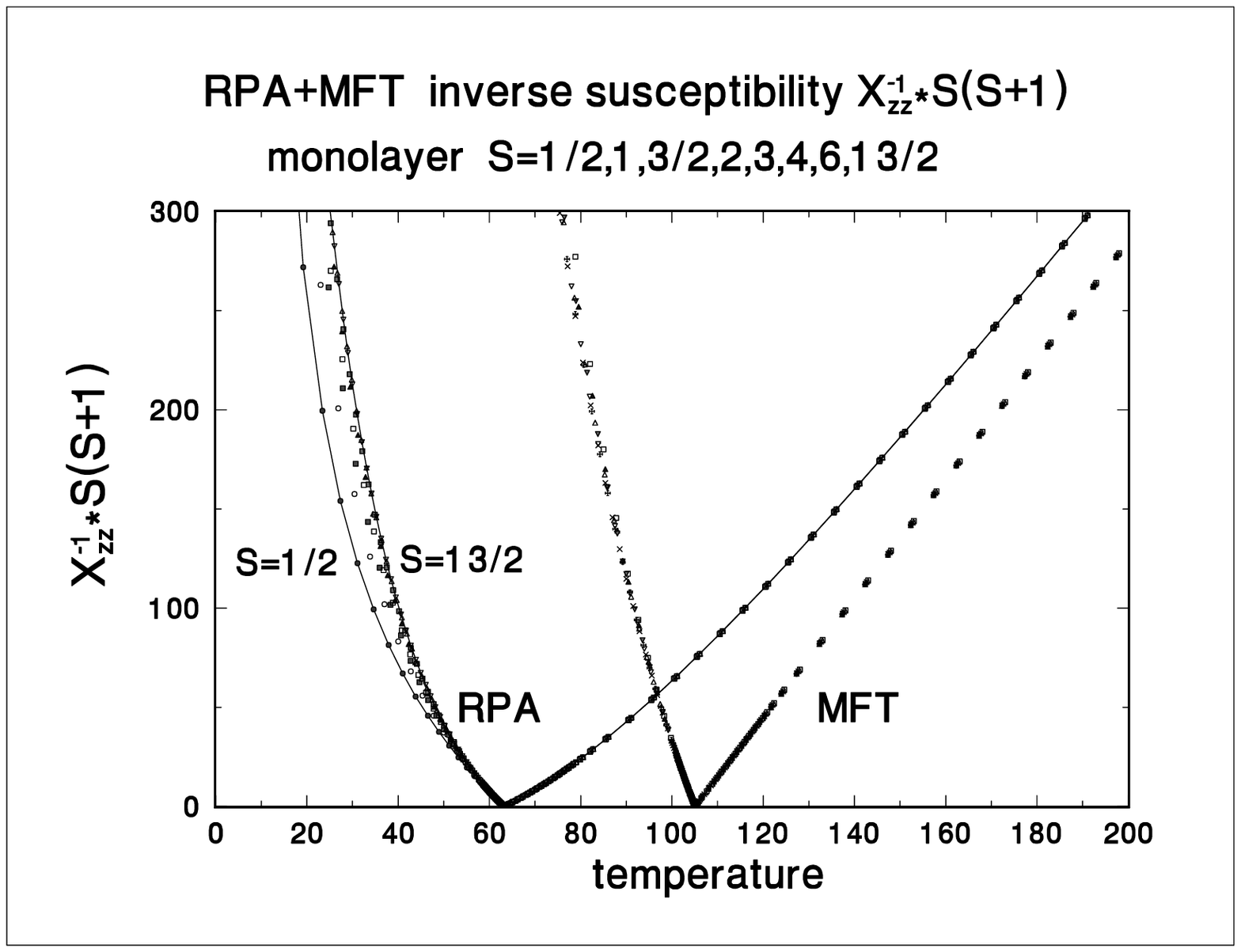}
\protect
\caption{ `Universal' inverse susceptibilities $\chi_{zz}^{-1}*S(S+1)$ along
the easy axis of an
anisotropic ferromagnetic Heisenberg monolayer for a square lattice
as functions of the temperature for $S=1/2,1,3/2,2,3.4,6,13/2$. Comparison
is made between Green's function (RPA) and mean field (MFT) calculations.}
\label{fig2}
\end{center}
\end{figure}
\begin{figure}[htb]
\begin{center}
\protect
\includegraphics*[bb = 80  90 520 420,
angle=-0,clip=true,width=9cm]{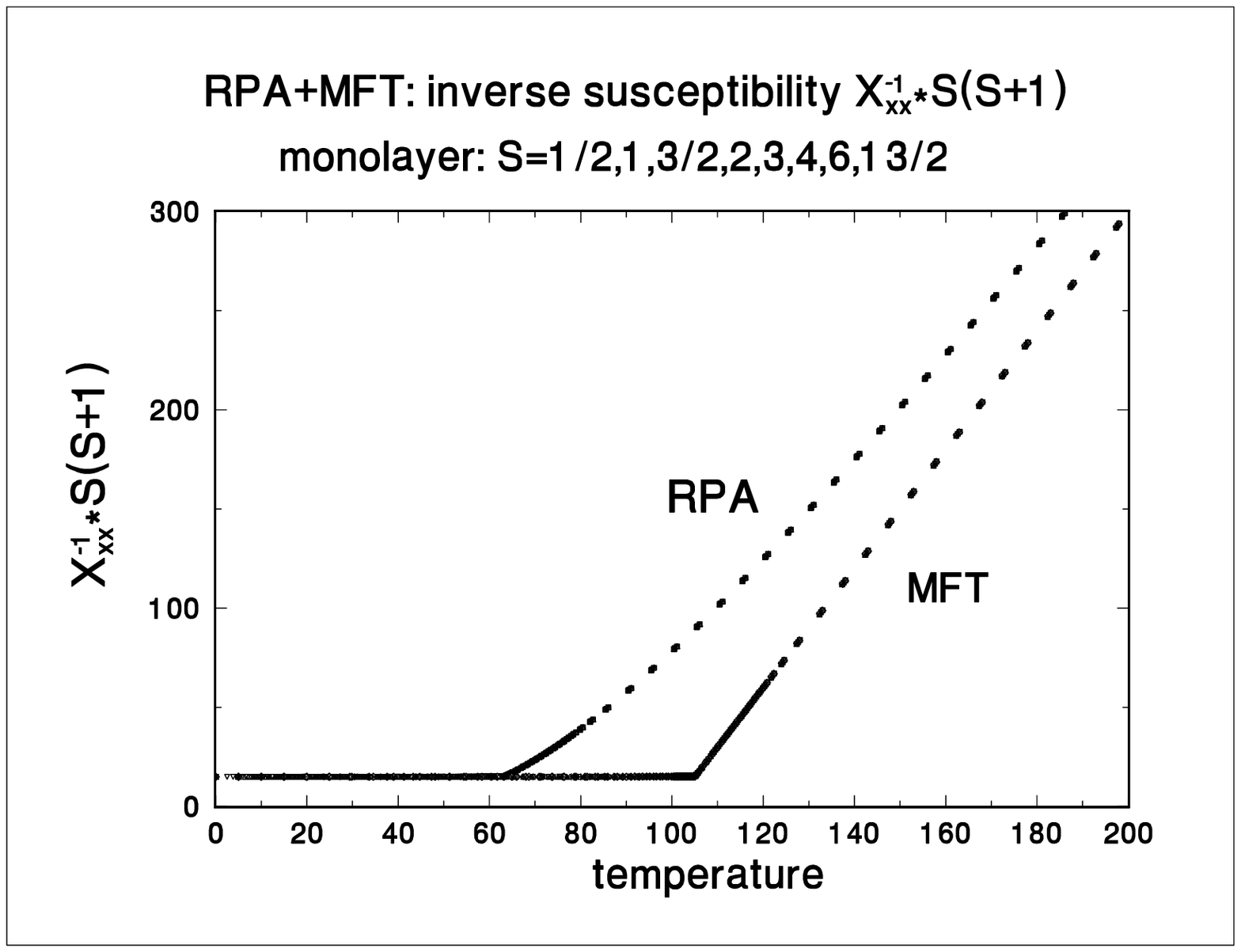}
\protect
\caption{The `universal' inverse susceptibilities $\chi_{xx}^{-1}*S(S+1)$ along
the hard axis of an
anisotropic ferromagnetic Heisenberg monolayer for a square lattice are shown
as functions of the temperature for $S=1/2,1,3/2,2,3,4,6,13/2$. Comparison
is made between the results of Green's function (RPA) and mean field (MFT)
calculations.}
\label{fig3}
\end{center}
\end{figure}
In Fig.2 we compare RPA and MFT calculations for the inverse susceptibility
$\chi_{zz}^{-1}S(S+1)$ along the easy axis. As in Fig. 1, there is a shift
in the Curie temperatures in going from RPA to MFT. For $T<T_C$,  the MFT
result behaves more universally
than that of RPA; above the Curie temperature, both results are nearly
universal.  $\chi_{zz}^{-1(MFT)}*S(S+1)$ is a straight line
$\propto (T-T_C^{MFT})$ like a Curie-Weiss law. For $S=1/2$, one finds
analytically from equation (\ref{30}) in the limit
$\la S^z \ra\rightarrow 0, T \rightarrow{\rm large}$, that
\begin{equation}
S(S+1)\chi_{zz}^{-1(MFT)}=\frac{3}{4}4(T-T_C^{MFT}),
\end{equation}
where $T_C^{MFT}=J+D$.
The
inverse RPA susceptibility $\chi_{zz}^{-1(RPA)}*S(S+1)$ is curved for
$T>T_C^{RPA}$
due to magnon effects. This is a behaviour known from isotropic bulk
ferromagnets, but the effect is significantly stronger for a monolayer.

An analogous universality is obtained for the inverse suceptibility
$\chi_{xx}^{-1}*S(S+1)$. The results are shown in Fig.3. Contrary to
the curve for $\chi_{zz}^{-1}$, the hard axis susceptibility
does not go to zero at $T=T_C$.
For $T<T_C$ one has the same universal constant in RPA and MFT, which
can be calculated analytically for the monolayer
$\chi_{xx}^{-1}*S(S+1)=\tilde{D}q$ from equation (\ref{32}). The
slopes of the curves for $T>T_C$ are, however, different. The MFT
again yields a straight line,
whereas the RPA result is curved and approaches a straight line only
for very large T. Hence, we see that owing
to the scaling properties, it is not necessary to do calculations for each
spin value separately. It suffices to do calculations for one spin value
and then to apply scaling to obtain the results for other spin values.

\subsection*{3.2. Multilayers at fixed spin S}

Next we discuss the multilayer case for fixed spin. We use the example of spin
$S=1/2$ (We have also considered multilayers with spins $S>1/2$.
The results
scale with respect to the spin as in the monolayer case).

Curie temperatures as function of the layer thickness are shown in Fig.4. The
difference between RPA and MFT is largest for the monolayer, where
$T^{RPA}_C\simeq 0.60\ T^{MFT}_C$ and shrinks to
$T^{RPA}_C\simeq 0.80\ T^{MFT}_C$
for a film with N=19 layers, where one is approaching the bulk value.
\begin{figure}[htp]
\begin{center}
\protect
\includegraphics*[bb = 80  90 520 420,
angle=-0,clip=true,width=9cm]{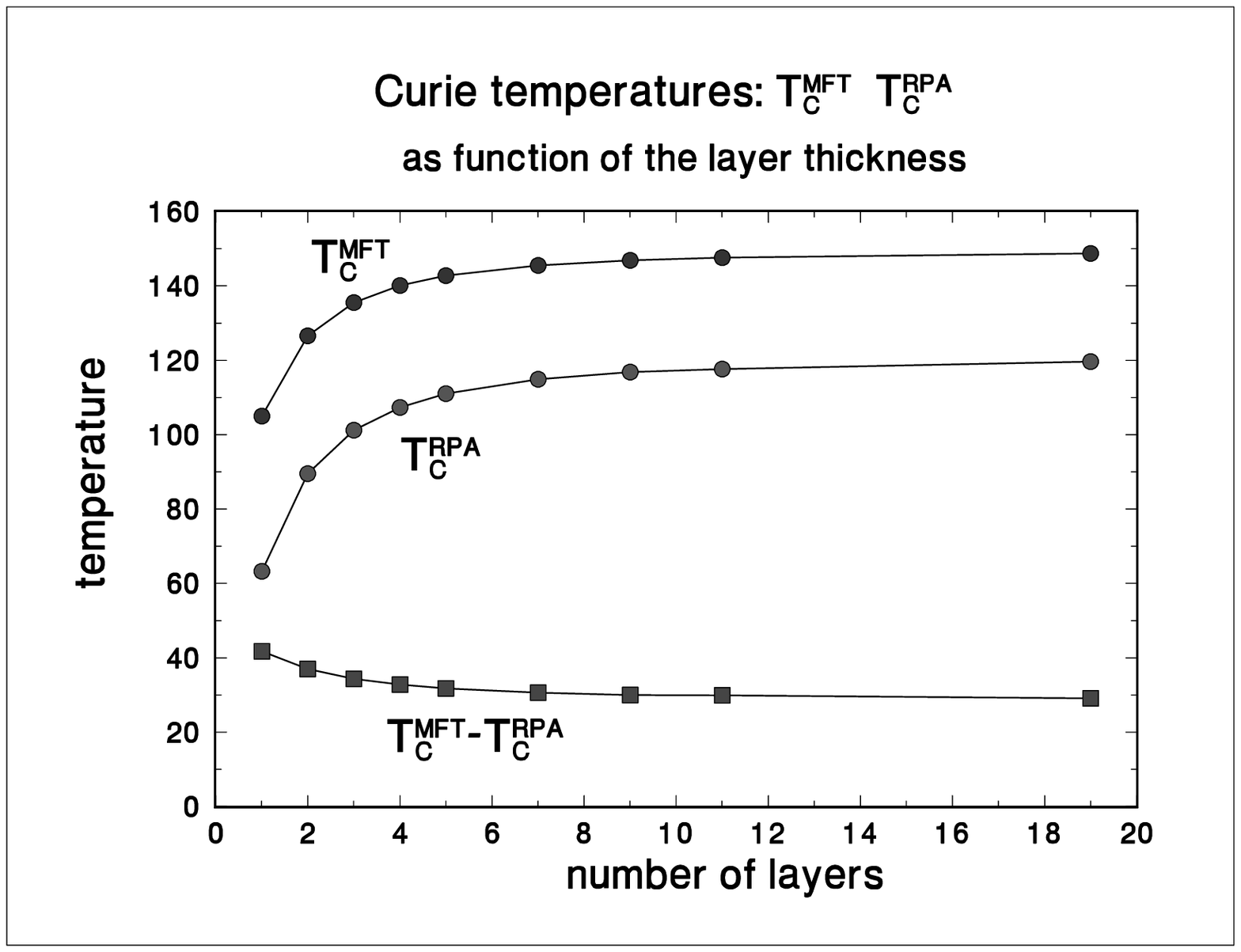}
\protect
\caption{
Curie temperatures of ferromagnetic films with spin $S=1/2$
are shown as functions of the film thickness for MFT ($T_C^{MFT}$) and RPA
($T_C^{RPA}$). The difference $T_C^{MFT}-T_C^{RPA}$ shrinks by a about a factor
of two when going from the monolayer to the bulk limit.}
\label{fig4}
\end{center}
\end{figure}

To further emphasize the difference between RPA and MFT, we show
in Figs. 5 and 6 the inverse susceptibilities $\chi_{zz}^{-1}$ and
$\chi_{xx}^{-1}$ as functions of the temperature. To avoid cluttering the
figures, we plot only the results of the monolayer (N=1) and the film with
the maximum number of layers (N=19), which is close to the
bulk limit
because the Curie temperatures saturate with increasing number of layers N.
In each case, we observe the shift in the Curie temperatures
between RPA and MFT
corresponding to Fig.4.
\begin{figure}[htp]
\begin{center}
\protect
\includegraphics*[bb = 80  90 520 420,
angle=-0,clip=true,width=9cm]{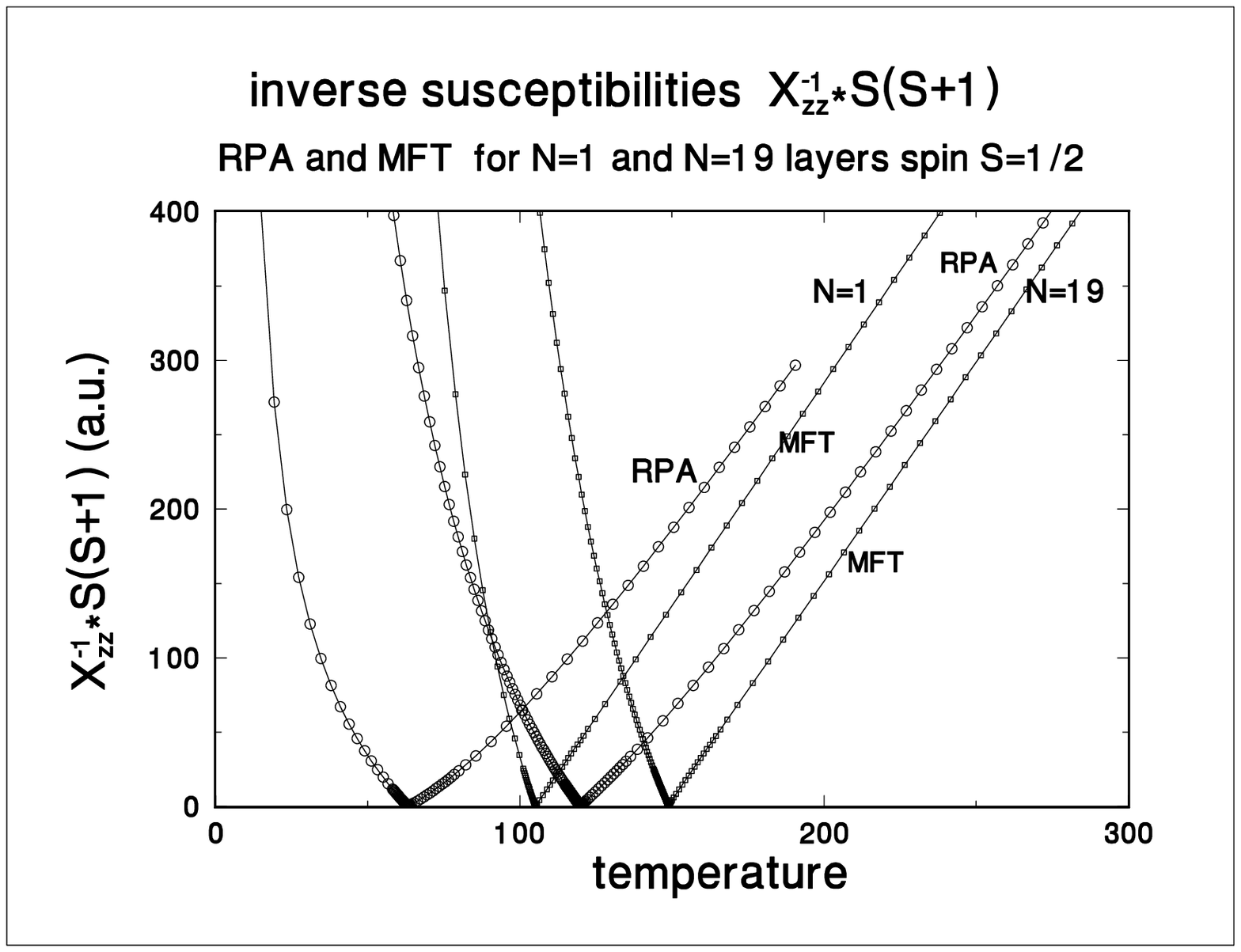}
\protect
\caption{The inverse susceptibilities $\chi_{zz}^{-1}*S(S+1)$ along the easy
axisof a
ferromagnetic film with spin $S=1/2$ for RPA and MFT are shown as a function of
the temperature for a monolayer (N=1) and a multilayer (N=19). }
\label{fig5}
\end{center}
\end{figure}
\begin{figure}[htp]
\begin{center}
\protect
\includegraphics*[bb = 80  90 520 420,
angle=-0,clip=true,width=9cm]{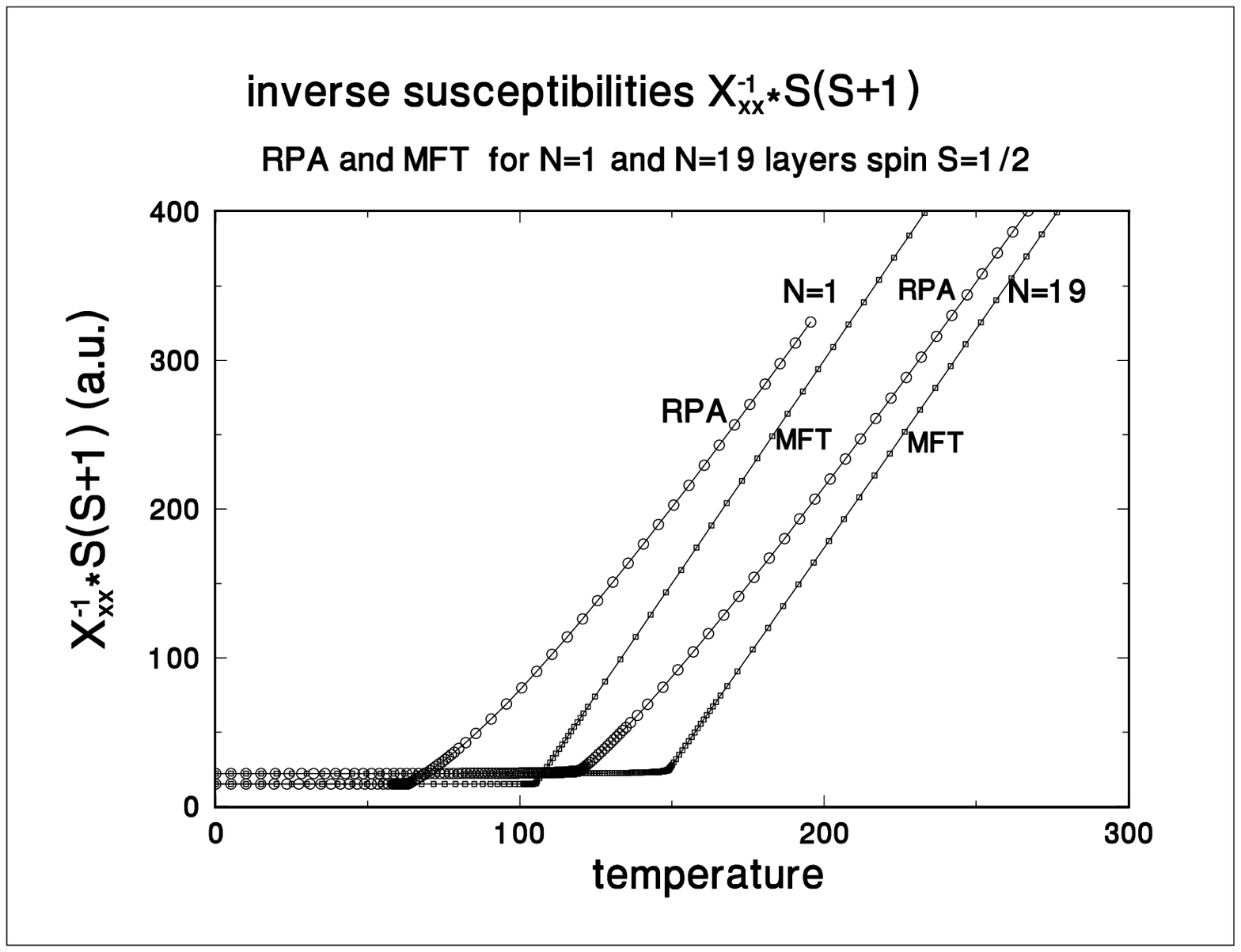}
\protect
\caption{The inverse susceptibilities $\chi_{xx}^{-1}*S(S+1)$ along the hard
axis of a
ferromagnetic film with spin $S=1/2$ for RPA and MFT are shown as a function of
the temperature for a monolayer (N=1) and a multilayer (N=19). }
\label{fig6}
\end{center}
\end{figure}

Above the Curie temperature,
the inverse susceptibilities are straight lines in the mean field case
and curved lines in RPA.
The slopes of the curves, however, are different for each film
thickness.
This is seen most clearly if one normalizes the temperature scale to the Curie
temperatures $T_C(N)$. The slopes in MFT increase with increasing film
thickness, and in RPA the curvature decreases with increasing number of
layers, as shown in fig.7.

For $T\leq T_C$ the inverse susceptibility $\chi_{xx}^{-1}$ is constant,
having the same value in MFT and RPA but depending on the number of layers.
The reason for the layer-dependence is that, with
increasing film thickness,
the number of nearest neighbours increases and therefore one has an increase in
the inverse susceptibility. For the square lattice monolayer and bilayer,
the values can
be calculated analytically from the regularity condition (\ref{19}):
$\chi_{xx}^{-1}(N=1)=\tilde{D}q$ and $\chi_{xx}^{-1}(N=2)=\tilde{D}(q+1)$,
with $q=4$ for a square lattice. From the value of $\chi_{xx}^{-1}$
at $T=T_C$, one can obtain an estimate
of the exchange anisotropy strength parameter D, which, together with the Curie
temperature, which depends on the exchange interaction strength J
and on the exchange
anisotropy strength D, affords an estimation of J.
As the number of layers $N$ increases, the relative weights of the
layers having two neighbouring layers increases, since it is only
the sites in the surface layers which are restricted to having one
nearest neighbour in the next layer; hence
$\chi_{xx}^{-1}$ slowly increases (see Fig. 8).
\begin{figure}[htb]
\begin{center}
\protect
\includegraphics*[bb = 80  90 520 420,
angle=-0,clip=true,width=9cm]{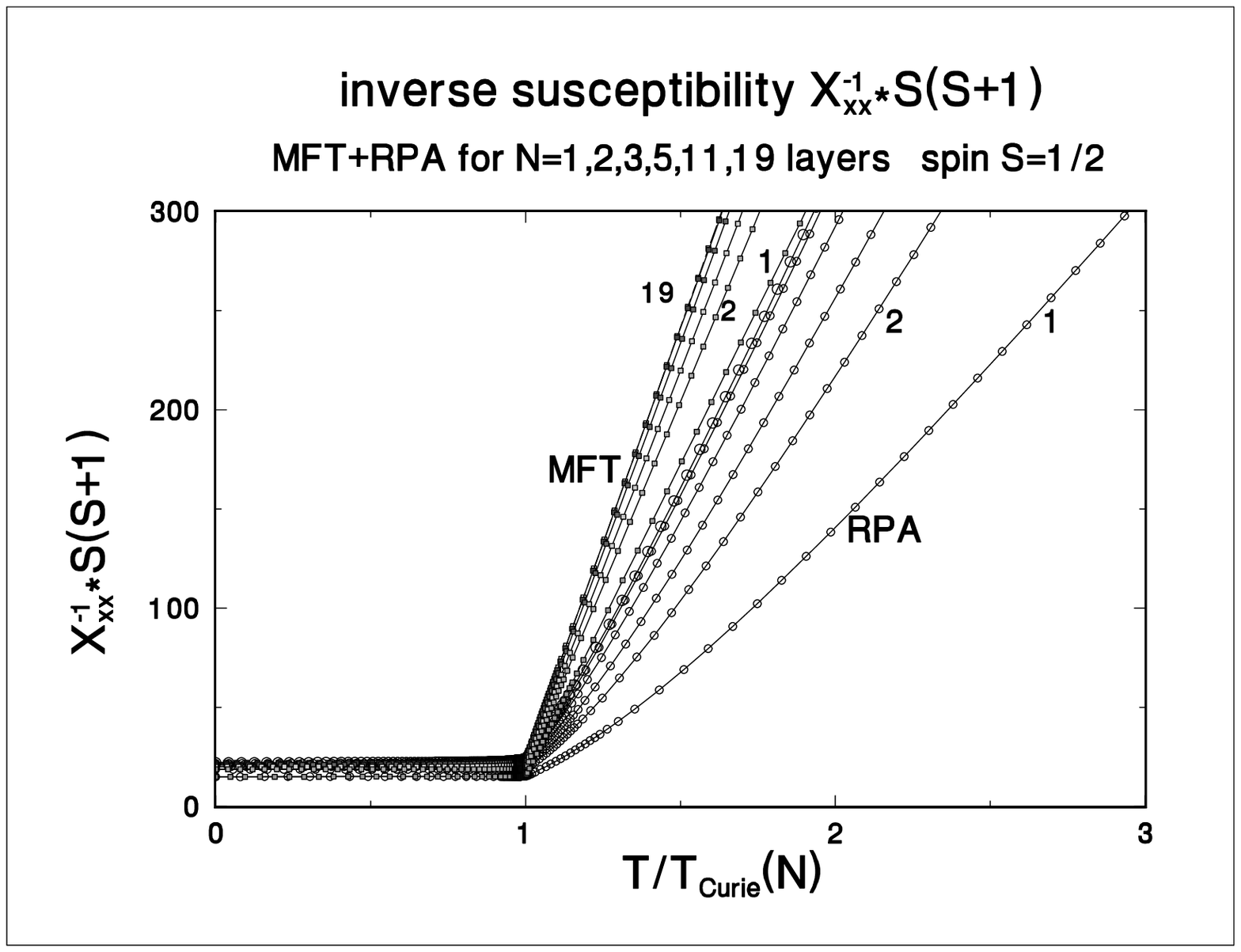}
\protect
\caption{The inverse susceptibilities $\chi_{xx}^{-1}*S(S+1)$ along the hard
axis of a
ferromagnetic film with spin $S=1/2$ for RPA and MFT are shown as a function of
the reduced temperature $T/T_{Curie}(N)$ for N=1,2,3,5,11,19 layers. }
\label{fig7}
\end{center}
\end{figure}
\begin{figure}[htb]
\begin{center}
\protect
\includegraphics*[bb = 80  90 520 410,
angle=-0,clip=true,width=9.cm]{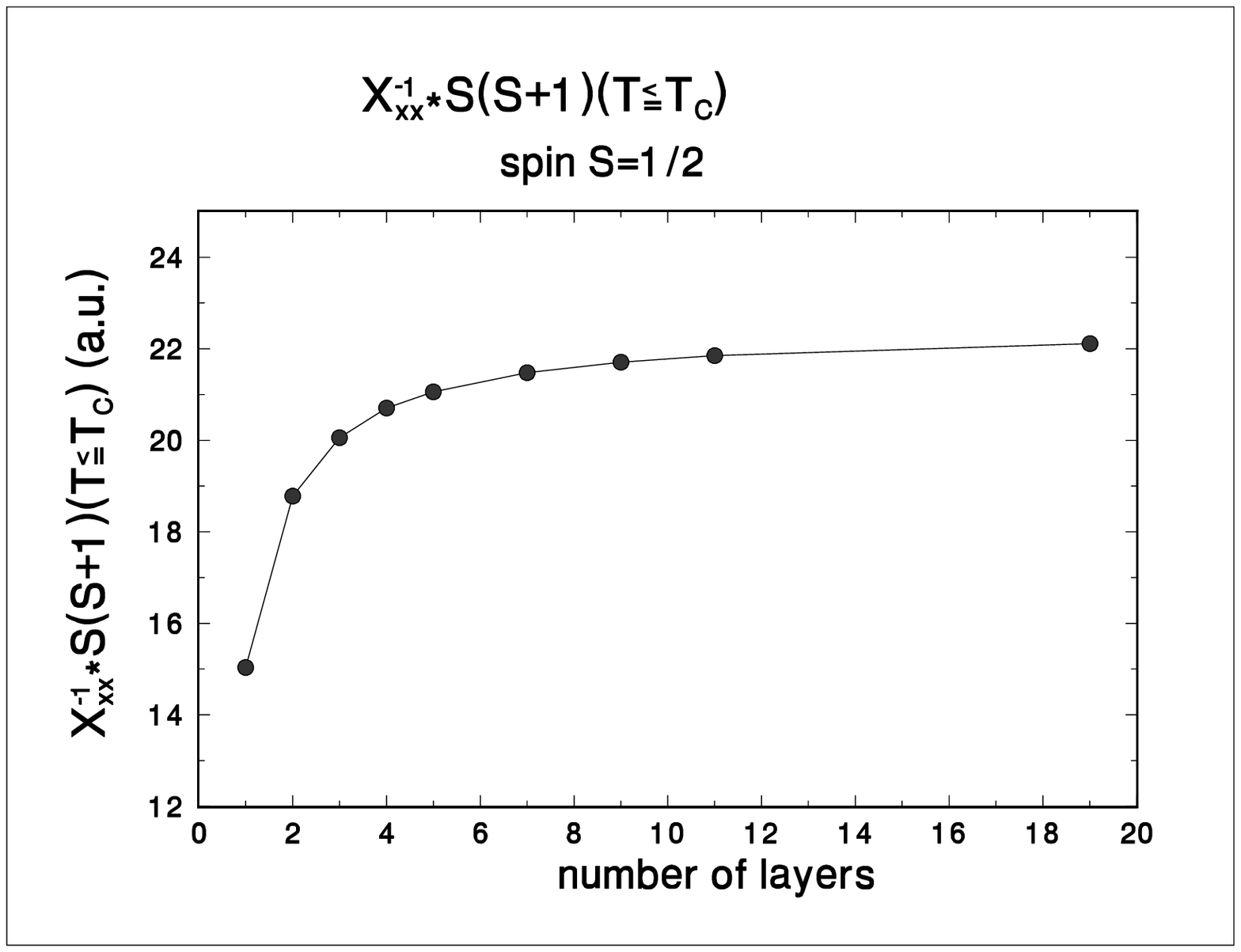}
\protect
\caption{The inverse susceptibilities $\chi_{xx}^{-1}*S(S+1)$ along the hard
axis at the Curie temperatures of a
ferromagnetic film with spin $S=1/2$ for RPA are shown as a function
of the film thickness. }
\label{fig8}
\end{center}
\end{figure}
\subsection*{4. Summary and conclusion}

We have generalized the many-body Green's function
treatment for calculating in-plane anisotropic static susceptibilities of
ferromagnetic films to arbitrary spin $S$ and to multilayers.
In particular, we have emphasized the difference in the results from a
Green's function theory (RPA) and mean field theory (MFT),
pointing out the importance of
spin waves, which are absent in MFT. All results discussed below refer to a
simple cubic lattice.

By introducing scaled variables, we were able to show that
the magnetic properties of
thin ferromagnetic films with in-plane anisotropy manifest nearly
universal behaviour.
Plotting $\la S^z\ra/S$ and $\la S^x\ra/(S+1)$ as functions of the temperature
reveals a nearly universal behaviour over the whole temperature
range for RPA, whereas
$\la S^z\ra/S$ for MFT shows a small dependence on S. The main
difference between
RPA and MFT is in the universal Curie temperature which, for the monolayer, is
nearly a factor of two larger in MFT than in RPA,
$T_C^{MFT}\simeq 2*T_C^{RPA}$, due to spin wave effects. The hard-axis
magnetization $\la S^x\ra$ has the same constant value in RPA and MFT for
$T\leq T_C$ because
it depends only on the number of nearest neighbours and not on the momentum
of the lattice.
The inverse susceptibilities along the easy and hard axes also behave
universally when scaled as $\chi_{zz}^{-1}*S(S+1)$ and $\chi_{xx}^{-1}*S(S+1)$,
particularly in the paramagnetic regime. The difference between MFT and RPA
consists in the shift of the Curie temperatures and the behaviour in the
paramagnetic region ($T>T_C$). Whereas the MFT inverse susceptibilities are
linear in $(T-T_C^{MFT})$ (a Curie-Weiss-like behaviour), the RPA
susceptibilities are curved, owing to spin-wave
effects, and approach a straight line only asymptotically for very large
temperatures. As long as one scales with respect to the spin  $S$ there is
no qualitative change
in the physical picture from that discussed in
Ref. \cite{Jens03} for spin $S=1/2$.
It is not necessary to perform calculations for each spin value separately.
Instead, it is sufficient to calculate the results for one spin value and to
apply scaling. This is one of the main results of the paper.

For the multilayers at a fixed spin $S$, the Curie
temperature increases with increasing film thickness, approaching the bulk
limit
around a film consisting of $N\simeq 19$ layers. The difference between the
Curie temperatures for MFT and RPA decreases with increasing film thickness from
$T_C^{MFT}/T_C^{RPA}(N=1)\simeq 2$ to $T_C^{MFT}/T_C^{RPA}(N=1)\simeq 1.3$,
which shows that the spin wave effects are strongest for the monolayer.
In MFT, the inverse susceptibilities show a
linear Curie-Weiss behaviour for $T>T_C$, whereas the RPA results are curved.
When plotting the inverse susceptibilities as a function of
the normalized temperatures $T/T_C(N)$
the slopes of the straight lines of MFT increase with increasing
layer
number, whereas  the curvatures of RPA decrease with increasing film thickness.
From the curvatures of the inverse
susceptibilities it is thus possible to deduce the number of layers, which
might be a way to extract information on the film thickness from experiment.
For $T\leq T_C$, the inverse hard axis susceptibilites $\chi_{xx}^{-1}$ are
constants,
but their value increases with increasing layer thickness, although not very
strongly, which
allows one to discriminate between films with differing numbers of layers.
From $\chi_{xx}^{-1}(T\leq T_C)$, one can get in principle information about
the exchange
anisotropy strength, whereas the value of the Curie temperature depends on
both the exchange interaction and the exchange anisotropy strengths.
All the results discussed might be modified by a layer-dependence
of the exchange interaction and exchange anisotropy or, when
a different lattice type has to be considered. Other effects like domain wall
motion or vortex excitations, which are not treated
in the theory above, could also lead to modifications.

The calculations here demonstrate that we are technically able to
calculate the magnetic properties of in-plane anisotropic ferromagnetic
multilayer films with $S\geq 1/2$. Hopefully, some of the predictions of the
present calculations can be verified experimentally in the future, in
particular with respect to the hard axis susceptibility. This might be possible
if the experimental techniques discussed in Ref. \cite{Jens03}, where
experimental results on a bilayer are reported, can be improved.
High precision measurements of anisotropic susceptibilities of thin films,
particularly above $T_C$, are called for, which is certainly a challenge for
experimentalists.

We are indebted to P.J. Jensen for useful discussions.

\end{document}